\documentclass[10pt,conference]{IEEEtran}
\IEEEoverridecommandlockouts

\usepackage{cite}
\usepackage{amsmath,amssymb,amsfonts}
\usepackage{algorithmic}
\usepackage{graphicx}
\usepackage{subfigure}
\usepackage{textcomp}
\usepackage{xcolor}
\usepackage{mdframed}
\usepackage{multirow}
\usepackage{tabularx}
\usepackage{listings}
\usepackage{booktabs}
\usepackage{amsthm}
\newtheorem{example}{Example}
\def\BibTeX{{\rm B\kern-.05em{\sc i\kern-.025em b}\kern-.08em
    T\kern-.1667em\lower.7ex\hbox{E}\kern-.125emX}}

\begin{document}

\title{When Abstraction Breaks Physics:\\ Rethinking Modular Design in Quantum Software}

 \author{\IEEEauthorblockN{Jianjun Zhao}
 \IEEEauthorblockA{\textit{Kyushu University, Japan} \\
zhao@ait.kyushu-u.ac.jp
}
}

\maketitle
\begin{abstract}
Abstraction is a fundamental principle in classical software engineering, which enables modularity, reusability, and scalability. However, quantum programs adhere to fundamentally different semantics, such as unitarity, entanglement, the no-cloning theorem, and the destructive nature of measurement, which introduce challenges to the safe use of classical abstraction mechanisms. This paper identifies a fundamental conflict in quantum software engineering: \emph{abstraction practices that are syntactically valid may violate the physical constraints of quantum computation}. We present three classes of failure cases where naive abstraction breaks quantum semantics and propose a set of design principles for physically sound abstraction mechanisms. We further propose research directions, including quantum-specific type systems, effect annotations, and contract-based module design. Our goal is to initiate a systematic rethinking of abstraction in quantum software engineering, based on quantum semantics and considering engineering scalability.
\end{abstract}


\section{Introduction}

Abstraction is a fundamental principle in classical software engineering~\cite{sommerville2011software}. By hiding low-level implementation details and exposing high-level interfaces, abstraction enables modularity, reusability, and scalability, which are key elements for building complex systems~\cite{parnas1972,shaw1990}. In classical programs, abstraction mechanisms such as functions, classes, and modules operate under assumptions of determinism, full observability, and compositional semantics~\cite{ghezzi2003,meyer1997object}. However, in the domain of quantum software engineering (QSE)~\cite{zhao2020qse}, these assumptions do not hold.

Quantum programs are governed by fundamentally different physical principles, including unitarity, entanglement, the no-cloning theorem, and the destructive nature of measurement~\cite{nielsen2010quantum}. These properties introduce key challenges to the safe use of classical abstraction mechanisms. For example, encapsulating a subroutine composed of quantum gates may unintentionally violate global unitarity. Similarly, abstracting away the entangled states across modules may disrupt quantum correlations, leading to incorrect or non-executable behavior. Measurement operations, which collapse quantum states irreversibly, cannot be abstracted like classical observations without semantic misrepresentation.

Despite growing interest in high-level quantum programming languages and modular development frameworks~\cite{green2013quipper,svore2018q,bichsel2020silq}, the foundational conflict between abstraction and quantum physical constraints has received little attention. Most existing languages focus on syntax or high-level constructs, but leave unresolved the question of whether such abstractions remain physically valid and semantically meaningful.
In this paper, we argue that the concept of abstraction in quantum programs must be rethought under the lens of quantum semantics. We identify key failure cases where naive abstraction leads to physically invalid or semantically misleading behavior. Based on these observations, we propose a set of design principles for \emph{physically sound abstraction mechanisms} that preserve unitarity, respect the boundaries of the entanglement, and account for the effects of the measurement. Our goal is to initiate a systematic rethinking of abstraction in QSE, laying the foundation for future language designs, toolchains, and modular development approaches.

\begin{table*}[t]
\small
\center{
\caption{Contrasting classical and quantum abstraction assumptions}
\label{tab:abstraction-differences}
\renewcommand{\arraystretch}{1.5} 
\setlength{\tabcolsep}{4pt}       
\begin{tabular}{|p{3cm}|p{5.8cm}|p{8.2cm}|}
\hline
\textbf{Aspect} & \textbf{Classical Programs} & \textbf{Quantum Programs} \\
\hline
State representation & Concrete values (e.g., integers, booleans) & Superpositions of basis states in a Hilbert space \\
\hline
Observability & Full inspection of intermediate states & Intermediate states are not observable due to collapse \\
\hline
Determinism & Execution is deterministic & Execution is inherently probabilistic \\
\hline
Compositionality & Modules can be safely composed & Composition may break unitarity or entanglement integrity
\\
\hline
Copying of data & Data can be freely copied & Qubits cannot be cloned due to the no-cloning theorem
\\
\hline
Measurement effects & Observations are non-destructive & Measurement irreversibly collapses the quantum state
\\
\hline
\end{tabular}
}
\end{table*}

\section{Why Classical Abstraction Fails in Quantum Programs}

Abstraction in classical software is enabled by assumptions that align with the semantics of deterministic computation. When a function is abstracted, its internal control flow and data manipulations are hidden, but its behavior remains predictable, reproducible, and compositionally safe. These abstractions facilitate reasoning, modularization, and code reuse, which form the basis of modern software engineering~\cite{ghezzi2003}.
However, when these classical abstraction mechanisms are applied directly to quantum programs, they often fail due to fundamental mismatches between the semantics of classical computation and quantum mechanics. Table~\ref{tab:abstraction-differences} summarizes the key differences between classical and quantum abstractions.
These differences show that classical abstractions implicitly rely on properties such as observability, reusability, and determinism, which are not valid in quantum computation. For example, abstracting a quantum subroutine without considering its entanglement with external qubits can result in incorrect program behavior. Similarly, encapsulating measurement operations without making their side effects explicit violates the semantics of quantum state evolution.
Moreover, classical abstraction assumes that modules can be composed without a global context. However, in quantum systems, whether a unitary transformation is valid often depends on the state of the entire system entangled, which makes local reasoning in general invalid.

In summary, the fundamental principles that form the basis of abstraction in classical software, such as safe composition, internal observability, and data duplication, are incompatible with the fundamental nature of quantum programs. These conflicts suggest that a physically sound notion of abstraction for quantum software must be based on the semantics of quantum mechanics, rather than classical computation.

\section{Failure Cases of Naive Abstraction}
\label{sec:failure}

While abstraction is a desirable engineering mechanism, applying it naively in quantum programs can result in physically invalid or semantically incorrect behavior. This section presents three representative failure cases observed in common quantum programming practices, each illustrating a fundamental conflict between abstraction and quantum semantics.

\subsubsection{Violation of Unitarity}

In classical programming, subroutines can encapsulate arbitrary computations. However, in quantum computing, unless measurement is involved, every operation must be unitary~\cite{nielsen2010quantum}. Encapsulating a non-unitary sequence, for example, conditionally applying gates based on classical logic, without proper control or ancillary handling, can result in an invalid transformation.

{\small
\begin{lstlisting}[language=Python, caption={Encapsulated function may break unitarity (conceptual example)}, label={lst:unitarity-break}]

  def my_subroutine(circuit, q):
      if some_classical_condition():
          circuit.x(q[0])
          circuit.h(q[1])
\end{lstlisting}
}

In Listing~\ref{lst:unitarity-break}, the subroutine \texttt{my\_subroutine} encapsulates a classically controlled quantum operation. While syntactically valid in Python, such patterns risk violating global unitarity when classical control is evaluated outside the context of the quantum circuit. This breaks the assumption that all quantum subcircuits must be expressed as unitary transformations, unless they explicitly involve measurement or reset.

\subsubsection{Entanglement Boundary Violation}

Quantum entanglement leads to non-local dependencies between qubits. Abstracting a subcircuit that operates on an entangled subset of qubits without explicitly modeling the entanglement boundary can result in hidden dependencies or unintended side effects when the module is reused or rearranged~\cite{yu2025dependence}.

{\small
\begin{lstlisting}[language=Python, caption={Abstracted entanglement operation reused across contexts}, label={lst:entanglement-break}]

  def entangle_pair(circuit, q0, q1):
      circuit.h(q0)
      circuit.cx(q0, q1)

  # Used in global context
  entangle_pair(circ, q[0], q[1])
  entangle_pair(circ, q[2], q[3])
\end{lstlisting}
}

In Listing~\ref{lst:entanglement-break}, the abstraction \texttt{entangle\_pair} is reused without awareness of the global entanglement structure. If other parts of the program also entangle \texttt{q1} or \texttt{q2} with other qubits, reusing this module may interfere with entanglement across modules and break the assumption that modules operate independently.

\subsubsection{Violation of Measurement Semantics}

Measurement in quantum computing is non-reversible and collapses quantum states~\cite{nielsen2010quantum}. Encapsulating measurements inside an abstract module may hide their global impact, especially when used within larger circuits or hybrid quantum-classical loops.

{\small
\begin{lstlisting}[language=Python, caption={Measurement hidden inside abstraction}, label={lst:measurement-break}]

  def measure_qubit(circuit, q, c, idx):
      circuit.h(q[idx])
      circuit.measure(q[idx], c[idx])
\end{lstlisting}
}

In Listing~\ref{lst:measurement-break}, the function \texttt{measure\_qubit} performs a Hadamard transform followed by measurement on a single qubit. The use of encapsulation hides the non-reversible nature of measurement from the calling context, potentially invalidating the assumptions of later circuit segments. From the outside, the measurement side-effect is not evident, which violates compositional transparency.

In summary, these failure cases demonstrate that abstraction in quantum software must not only preserve syntactic validity but also maintain physical correctness. In the next section, we identify core design principles for abstraction mechanisms that are compatible with quantum semantics.

\section{Design Principles for Physically Sound Abstractions}

To handle the conflicts discussed in Section~\ref{sec:failure}, we propose several design principles for abstraction mechanisms in quantum software. These principles aim to ensure that abstractions preserve the physical semantics of quantum computation and remain valid across modular compositions. Unlike classical abstraction guidelines that primarily target structural correctness and interface clarity, these principles are based on quantum physics and semantics.

\subsubsection{Preservation of Unitarity}

Any abstracted quantum subroutine must either (i) be representable as a unitary transformation over the involved qubits, or (ii) explicitly declare the presence of non-unitary effects such as measurement or reset. This requirement ensures that the composition of abstracted modules does not implicitly violate the core constraint of quantum evolution.

\begin{mdframed}[backgroundcolor=gray!10, linecolor=black, roundcorner=20pt, skipabove=10pt]
\noindent \textbf{Implication}: Language-level abstractions (e.g., circuit functions, macros) should enforce or annotate unitarity, and compilers should provide warnings if composed abstractions break global unitarity without explicit intent.
\end{mdframed}

\subsubsection{Entanglement Boundary Awareness}

Abstractions must not hide entanglement relationships between qubits. When a module operates on qubits that are (or may be) entangled with external qubits, such dependences~\cite{yu2025dependence} should be made explicit, either through type annotations, interface contracts, or analysis tools~\cite{xia2023static}.

\begin{mdframed}[backgroundcolor=gray!10, linecolor=black, roundcorner=20pt, skipabove=8pt]
\noindent \textbf{Implication}: Toolchains should support entanglement tracking across modules. Programmers should be aware that copying or reusing a subcircuit may unintentionally propagate or break entanglement structures.
\end{mdframed}

\subsubsection{Measurement Transparency}
Measurement is a destructive operation that fundamentally alters the quantum state~\cite{nielsen2010quantum}. Any abstraction that performs measurement must make this fact explicit in its interface or usage contract. Implicit measurement within an encapsulated routine may break composability and violate the programmer’s assumptions about the continuity of quantum state evolution.

\begin{mdframed}[backgroundcolor=gray!10, linecolor=black, roundcorner=20pt, skipabove=8pt]
{
\textbf{Implication}: Quantum domain-specific languages (DSLs) should distinguish between pure and measurement-containing modules, and runtime systems should prohibit implicit reuse of measurement abstractions in contexts that assume unitary evolution.
}
\end{mdframed}

\subsubsection{Interface-Driven Classical Interaction}
In hybrid quantum-classical programs, the classical control flow must be handled via well-defined interfaces rather than opaque side effects. Abstracting classical logic together with quantum operations can obscure program semantics and break the deferred execution models used by many backends.

\begin{mdframed}[backgroundcolor=gray!10, linecolor=black, roundcorner=20pt, skipabove=8pt]
{
\textbf{Implication}: Classical logic should be modularized separately, and abstractions should clearly separate quantum operations from classical conditionals or callbacks that may affect circuit generation.
}
\end{mdframed}

\subsubsection{Physical Resource Compatibility}

Abstract modules should not assume arbitrary qubit connectivity or reuse without regard to underlying hardware constraints. Since physical qubit layout and coherence times affect execution fidelity, abstractions must be designed to either (i) remain hardware-agnostic and compilable, or (ii) declare hardware assumptions explicitly~\cite{shi2020resource}.

\begin{mdframed}[backgroundcolor=gray!10, linecolor=black, roundcorner=20pt, skipabove=8pt]
{
\textbf{Implication}: The extraction layers must support qubit allocation hints, routing awareness, or post-mapping validation to ensure compatibility of the real device.
}
\end{mdframed}
{
In summary, these principles provide a foundation for designing future quantum programming languages and toolchains that support modularity while preserving physical correctness. Rather than mimicking classical abstraction practices, quantum software engineering must develop its own abstraction discipline aligned with quantum mechanics.
}

\section{Rethinking Abstraction Mechanisms}

The need for physically sound abstraction mechanisms may open a new frontier in quantum software engineering, whereas classical software relies on decades of well-established design patterns and modeling formalisms~\cite{gamma1995design,harel1987statecharts,sommerville2011software}, quantum programming demands fundamentally different abstractions that are compatible with the laws of quantum mechanics. In this section, we present several research directions that could serve as a foundation for future quantum abstraction mechanisms.

\subsubsection{Type Systems for Abstraction Semantics} One promising direction is the development of quantum-specific type systems that encode abstraction constraints~\cite{yuan2022twist}. For example, a type system could statically enforce whether a module preserves unitarity, modifies entanglement, or performs measurement. These types would serve not only as documentation, but also as compile-time guarantees for physically valid behavior.

\begin{example}
A quantum subroutine \texttt{apply\_oracle} may be annotated with a type like \texttt{Unitary :: Qubit[n] -> Qubit[n]}, indicating that it is a unitary transformation that preserves quantum coherence. Conversely, a function like \texttt{measure\_qubits} may be typed as \texttt{Measuring :: Qubit[n] -> Bit[n]}, making it clear that it performs a destructive measurement and returns classical results.
\end{example}

\subsubsection{Effect Systems and Side-Effect Annotations} Quantum abstractions may benefit from effect annotations that track semantic side effects such as measurement, entanglement, and classical branching. These annotations can help developers reason about the behavior of modules without examining internal implementation and allow compilers or analyzers to catch violations of expected behavior.

\begin{example} A subroutine could be annotated with \texttt{@measuring}, indicating that it collapses part of the quantum state. Another might use \texttt{@entangles(q0, q1)} to indicate that it introduces entanglement between two qubits. These annotations would allow downstream tools to enforce constraints, such as forbidding the reuse of measured qubits in a coherent subcircuit.
\end{example}

\subsubsection{Contract-Based Module Design} 
Inspired by design-by-contract principles in classical software engineering~\cite{meyer1992applying}, quantum modules could specify preconditions and postconditions based on quantum mechanics, such as constraints on unitarity, entanglement, or measurement. These contracts would capture semantic constraints regarding the quantum state before and after execution, allowing tools or simulators to validate the correctness of usage.

\begin{example} A modular subroutine implementing a quantum kernel may declare that a contract like \texttt{requires: q[0] and q[1] are separable}, meaning that it expects the two qubits to be unentangled before invocation. Another example might specify \texttt{ensures: q[0] is disentangled after execution}, guiding a safe modular composition in larger quantum programs.
\end{example}

\subsubsection{Abstraction-Aware Tool Support}
The practicality of physically sound abstraction mechanisms depends on the corresponding tooling infrastructure. Tool support can provide feedback during development and debugging, especially when physical correctness is not visually obvious.

\begin{itemize}
  \item \textbf{Static analyzers}: Tools that scan quantum programs for entanglement leakage, unintended measurements, or violations of declared abstraction types~\cite{xia2023static}.
  \item \textbf{Visualization tools}: Interfaces that allow developers to inspect entanglement graphs and abstraction boundaries, improving their understanding of the program.
  \item \textbf{Verification frameworks}: Lightweight or formal tools to check whether a composed module preserves unitarity or whether all measurement effects are visible in abstraction contracts~\cite{alpay2020qwire}.
\end{itemize}

By embedding physical awareness into language constructs and toolchains, these approaches would allow developers to use abstraction safely, promoting modularity and scalability without compromising correctness.
As quantum programs grow in complexity, such mechanisms will be essential not only for reliability but also for collaboration, maintenance, and long-term evolution of quantum software systems.

\section{Related Work}

Recent quantum programming languages and frameworks emphasize modularity and scalability. Qiskit\cite{javadi2024quantum} supports modular circuit construction through Python functions and the \texttt{QuantumCircuit} class, but it does not provide checks for unitarity, entanglement boundaries, or measurement side effects, leaving these to the programmer.
Q\#\cite{svore2018q} introduces higher-level abstractions with adjoint and controlled operations, and separates classical from quantum data. However, it does not enforce rules for measurement visibility or entanglement safety, which can lead to misuse.
Silq\cite{bichsel2020silq} improves safety through automatic resource management and ancilla clean-up, but does not track measurement or entanglement, focusing mainly on resource safety. Quipper\cite{green2013quipper} generates scalable circuits using functional programming and parameterized abstractions, but its abstractions are mainly syntactic and do not validate physical correctness. ProjectQ~\cite{Steiger2018} and Cirq~\cite{Cirq2022} provide circuit modularization but lack mechanisms to enforce quantum semantics. Zayas Gallardo et al.~\cite{Zayas2025Locus} proposed \textit{Locus}, a programming abstraction for composing circuits with support for deferred instantiation and modular reuse. While their focus is on a concrete mechanism to improve compositionality in Qiskit, our work addresses the more fundamental issue of ensuring that abstraction mechanisms remain consistent with quantum semantics and physical constraints.

Several studies have explored modularity in quantum software, primarily through modeling rather than direct implementation. An early effort proposed a UML-based quantum software modeling language~\cite{perez2020towards}, which emphasized modularity. This was followed by several other modeling approaches~\cite{11126620,perez2022design,ali2020modeling}. However, these modeling languages do not adequately handle physical constraints such as unitarity and entanglement. A formal definition of quantum programming modules was proposed in~\cite{sanchez2021definition}, focusing on structural composition and encapsulation, but without considering quantum-specific concerns such as measurement collapse or entanglement preservation. Di Matteo~\cite{di2025art} further emphasized the general importance of abstraction in quantum software, but her discussion was limited to high-level perspectives and tool support rather than addressing semantic consistency with quantum physics.

To the best of our knowledge, no existing quantum programming language or framework systematically addresses physically sound abstraction, i.e., preserving unitarity, maintaining entanglement locality, and making measurement effects explicit. Our work is the first to identify this as a key challenge in quantum software engineering. Rather than proposing new language features, we argue that abstraction itself should be rethought in light of quantum constraints. This complements ongoing efforts in quantum language design, compiler optimization, and verification by focusing on an overlooked aspect: the correctness of abstraction under physical constraints.

\section{Conclusion and Future Work}

Abstraction is a fundamental principle in classical software engineering, but directly applying it to quantum programming faces key challenges. We argue that classical abstraction mechanisms based on determinism, observability, and compositionality are often incompatible with quantum physical semantics. We identify three types of failure cases where naive abstraction violates unitarity, fails to expose measurement, or disrupts entanglement. Based on these observations, we propose design principles for \emph{physically sound abstraction}, guided by the constraints of quantum mechanics. These include preserving unitarity, exposing measurement effects, maintaining entanglement locality, and aligning abstractions with hardware and classical interfaces.
For future work, we propose several directions for developing languages and tools that align with these principles. These include type systems that capture abstraction semantics, effect annotations for physical side effects, contract-based module design, and tool development for entanglement-aware validation and visualization.

As quantum software systems grow in size and complexity, the development of such abstraction mechanisms will be crucial to the reliability, reusability, and engineering scalability. We hope that this work could initiate a broader effort toward formalizing abstraction mechanisms that are both physically valid and practically useful in quantum software engineering.

\bibliographystyle{IEEEtran}
\bibliography{IEEEabrv}

@String{Computing = "Computing" }

@String{Computer = "{IEEE} Computer" }

@String{Springer = "Springer-Verlag" }

@article{javadi2024quantum,
  title={Quantum computing with Qiskit},
  author={Javadi-Abhari, Ali and Treinish, Matthew and Krsulich, Kevin and Wood, Christopher J and Lishman, Jake and Gacon, Julien and Martiel, Simon and Nation, Paul D and Bishop, Lev S and Cross, Andrew W and others},
  journal={arXiv preprint arXiv:2405.08810},
  year={2024}
}

@inproceedings{bichsel2020silq,
  title={Silq: A high-level quantum language with safe uncomputation and intuitive semantics},
  author={Bichsel, Benjamin and Baader, Maximilian and Gehr, Timon and Vechev, Martin},
  booktitle={Proceedings of the 41st ACM SIGPLAN Conference on Programming Language Design and Implementation},
  pages={286--300},
  year={2020}
}

@inproceedings{svore2018q,
  title={Q\# enabling scalable quantum computing and development with a high-level {DSL}},
  author={Svore, Krysta and Geller, Alan and Troyer, Matthias and Azariah, John and Granade, Christopher and Heim, Bettina and Kliuchnikov, Vadym and Mykhailova, Mariia and Paz, Andres and Roetteler, Martin},
  booktitle={Proceedings of the real world domain specific languages workshop 2018},
  pages={1--10},
  year={2018}
}

@article{parnas1972,
  author    = {David L. Parnas},
  title     = {On the Criteria To Be Used in Decomposing Systems into Modules},
  journal   = {Commun. ACM},
  volume    = {15},
  number    = {12},
  year      = {1972},
  pages     = {1053--1058}
}

@article{shaw1990,
  author    = {Mary Shaw},
  title     = {Prospects for an Engineering Discipline of Software},
  journal   = {IEEE Software},
  volume    = {7},
  number    = {6},
  pages     = {15--24},
  year      = {1990}
}

@book{ghezzi2003,
  author    = {Carlo Ghezzi and Mehdi Jazayeri and Dino Mandrioli},
  title     = {Fundamentals of Software Engineering},
  edition   = {2nd},
  publisher = {Prentice Hall},
  year      = {2003}
}

@book{nielsen2010quantum,
  title={Quantum computation and quantum information},
  author={Nielsen, Michael A and Chuang, Isaac L},
  year={2010},
  publisher={Cambridge university press}
}

@misc{zhao2020qse,
  author    = {Jianjun Zhao},
  title     = {Quantum Software Engineering: Landscapes and Horizons},
  howpublished = {arXiv preprint arXiv:2007.07047},
  year      = {2020}
}

@inproceedings{yu2025dependence,
  author    = {Haibo Yu and Jianjun Zhao},
  title     = {The Quantum Program Dependence Graph and Its Uses in Quantum Software Development},
  booktitle = {Proceedings of the International Workshop on Quantum Software Engineering (Q‑SE 2025)},
  year      = {2025},
  publisher = {IEEE/ACM},
  note      = {To appear}
}

@book{meyer1992applying,
  author    = {Bertrand Meyer},
  title     = {Applying ``Design by Contract''},
  year      = {1992},
  publisher = {Prentice Hall}
}

@inproceedings{green2013quipper,
  title={Quipper: a scalable quantum programming language},
  author={Green, Alexander S and Lumsdaine, Peter LeFanu and Ross, Neil J and Selinger, Peter and Valiron, Beno{\^\i}t},
  booktitle={Proceedings of the 34th ACM SIGPLAN conference on Programming language design and implementation},
  pages={333--342},
  year={2013}
}

@inproceedings{alpay2020qwire,
  author    = {Michael Alpay and Robert Rand and Steve Zdancewic},
  title     = {Q{WIRE}: Reasoning about quantum circuits},
  booktitle = {Proceedings of Quantum Physics and Logic (QPL)},
  year      = {2020},
  url       = {https://qpl2020.org/slides/qwire.pdf}
}

@inproceedings{xia2023static,
  title={Static entanglement analysis of quantum programs},
  author={Xia, Shangzhou and Zhao, Jianjun},
  booktitle={2023 IEEE/ACM 4th International Workshop on Quantum Software Engineering (Q-SE)},
  pages={42--49},
  year={2023},
  organization={IEEE}
}

@inproceedings{perez2020towards,
  title={Towards a quantum software modeling language},
  author={P{\'e}rez-Delgado, Carlos A and Perez-Gonzalez, Hector G},
  booktitle={Proceedings of the IEEE/ACM 42nd International Conference on Software Engineering Workshops},
  pages={442--444},
  year={2020}
}

@article{Steiger2018,
  author = {Steiger, Damian S. and Häner, Thomas and Troyer, Matthias},
  title = {Project{Q}: An Open Source Software Framework for Quantum Computing},
  journal = {Quantum},
  volume = {2},
  pages = {49},
  year = {2018},
  publisher = {Quantum Open Source Foundation}
}

@misc{Cirq2022,
  author = {Quantum AI team and collaborators},
  title = {Cirq},
  year = {2022},
  howpublished = {\url{https://github.com/quantumlib/Cirq}},
  note = {Accessed: 2024-06-11}
}

@book{gamma1995design,
  title={Design patterns: elements of reusable object-oriented software},
  author={Gamma, Erich and Helm, Richard and Johnson, Ralph and Vlissides, John},
  year={1995},
  publisher={Pearson Deutschland GmbH}
}

@article{harel1987statecharts,
  title={Statecharts: A visual formalism for complex systems},
  author={Harel, David},
  journal={Science of computer programming},
  volume={8},
  number={3},
  pages={231--274},
  year={1987},
  publisher={Elsevier}
}

@book{sommerville2011software,
  title={Software Engineering, 9/E},
  author={Sommerville, Ian},
  year={2011},
  publisher={America: Pearson Education Inc}
}

@book{meyer1997object,
  title={Object-oriented software construction},
  author={Meyer, Bertrand},
  volume={2},
  year={1997},
  publisher={Prentice hall Englewood Cliffs}
}

@article{shi2020resource,
  title={Resource-efficient quantum computing by breaking abstractions},
  author={Shi, Yunong and Gokhale, Pranav and Murali, Prakash and Baker, Jonathan M and Duckering, Casey and Ding, Yongshan and Brown, Natalie C and Chamberland, Christopher and Javadi-Abhari, Ali and Cross, Andrew W and others},
  journal={Proceedings of the IEEE},
  volume={108},
  number={8},
  pages={1353--1370},
  year={2020},
  publisher={IEEE}
}

@INPROCEEDINGS{11126620,
  author={Guo, Xiaoyu and Saito, Shinobu and Zhao, Jianjun},
  booktitle={2025 IEEE 49th Annual Computers, Software, and Applications Conference (COMPSAC)}, 
  title={Quan{UML}: Towards A Modeling Language for Model-Driven Quantum Software Development}, 
  year={2025},
  volume={},
  number={},
  pages={1344-1349},
  doi={10.1109/COMPSAC65507.2025.00168}}

@article{perez2022design,
  title={Design of classical-quantum systems with {UML}},
  author={P{\'e}rez-Castillo, Ricardo and Piattini, Mario},
  journal={Computing},
  volume={104},
  number={11},
  pages={2375--2403},
  year={2022},
  publisher={Springer}
}

@inproceedings{ali2020modeling,
  title={Modeling Quantum programs: challenges, initial results, and research directions},
  author={Ali, Shaukat and Yue, Tao},
  booktitle={Proceedings of the 1st ACM SIGSOFT International Workshop on Architectures and Paradigms for Engineering Quantum Software},
  pages={14--21},
  year={2020}
}

@article{sanchez2021definition,
  title={On the definition of quantum programming modules},
  author={S{\'a}nchez, Pedro and Alonso, Diego},
  journal={Applied Sciences},
  volume={11},
  number={13},
  pages={5843},
  year={2021},
  publisher={MDPI}
}

@inproceedings{zayas2025locus,
  title={Locus: A Proposal for Quantum Software Composition},
  author={Zayas Gallardo, Javier and Chicano, Francisco and Canal, Carlos and Murillo, Juan Manuel},
  booktitle={Companion Proceedings of the 9th International Conference on the Art, Science, and Engineering of Programming (Programming 2025)},
  pages={17--1},
  year={2025},
  organization={Schloss Dagstuhl--Leibniz-Zentrum f{\"u}r Informatik}
}

@inproceedings{di2025art,
  title={The Art of Abstraction in Quantum Software},
  author={Di Matteo, Olivia},
  booktitle={2025 IEEE/ACM International Workshop on Quantum Software Engineering (Q-SE)},
  pages={25--26},
  year={2025},
  organization={IEEE}
}

@article{yuan2022twist,
  title={Twist: Sound reasoning for purity and entanglement in quantum programs},
  author={Yuan, Charles and McNally, Christopher and Carbin, Michael},
  journal={Proceedings of the ACM on Programming Languages},
  volume={6},
  number={POPL},
  pages={1--32},
  year={2022},
  publisher={ACM New York, NY, USA}
}

\end{document}